# Dirac Bilayer Metasurfaces as an Inverse Gires-Tournois Etalon

Ki Young Lee[1,2], Kwang Wook Yoo[1], Francesco Monticone[3], and Jae Woong Yoon[1,2]*

[1]*Department of Physics, Hanyang University, Seoul 04763, Korea*
[2]*Research Institute for Natural Sciences, Hanyang University, Seoul 04763, Korea*
[3]*School of Electrical and Computer Engineering, Cornell University, Ithaca, NY 14853, United States.*
‡Correspondence should be addressed to yoonjw@hanyang.ac.kr

Efficient transmissive pure-phase resonances are highly desirable for optical modulation and wavefront engineering. Here, we propose a novel principle to realize a pure-phase resonance in an extremely broad transmission band, as opposed to previous approaches restricted to operating in reflection mode or over a narrow spectral band. We show that a glide-symmetric bilayer metasurface mathematically mimicking a two-dimensional Dirac semimetal induces unidirectional guided-mode excitation and perfect leakage-radiation blazing at the transmission channel. These effects create a peculiar resonant-scattering configuration, similar to the classical reflective Gires-Tournois etalon, but in transmission, providing full $2\pi$ phase modulation with constant transmittance near 100%. Most importantly, this effect persists over an extremely wide band, associated with topological effects. Hence, our proposed approach produces a spectrally and parametrically robust pure-phase resonance effect in transmission, which is highly beneficial for practical applications.

## I. INTRODUCTION

Nanophotonic phase-delay modulation enables various optical devices including high-speed phase shifters [1], solid-state beam scanners [2,3], 3D display pixels [4,5], spatial light modulators [6,7], orbital-angular-momentum generators [8], metasurfaces [9], and many others. In order to obtain remarkably large phase modulation from subwavelength thin-film structures, resonant excitation of localized states is typically required to produce a phase-delay range greatly enhanced from the non-resonant spectral background. Unfortunately, resonant phase delay is usually accompanied by abrupt intensity variations, which is detrimental for most applications.

A pure-phase resonance is a peculiar effect in which a phase delay can be realized over a full $2\pi$ range, as a function of certain parameters (e.g., frequency or structural parameters) while keeping the intensity constant, thereby providing an ideal mechanism for nanophotonic optical-phase control devices. A Gires-Tournois (GT) etalon is a well-known classical system that produces pure-phase resonances [10]. An ideal GT etalon consists of two parallel lossless mirrors, one is partially reflecting and the other is totally reflecting. The existence of a totally reflecting mirror makes the reflectivity of the system fixed at 100%, while the reflection phase delay across each interferometric resonance changes over a $2\pi$ range. An identical effect is obtainable in general if any lossless resonance mechanism is included in a totally reflecting optical system. Therefore, nanophotonic pure-phase resonance structures in such configurations have been widely investigated and developed so far [11–14]. However, the effect is produced only in reflection, causing various technical complications in practice, such as the need for additional redirecting optical paths, alignment issues, efficiency degradation, and optical window restriction.

For these reasons, transmissive pure-phase resonances are of great interest and some promising structures have been previously proposed in the literature. Two main principles have been investigated in this context. One is based on the combination of spectrally overlapped electric and magnetic dipole resonances in metasurface structures, i.e., periodic arrays of scatterers, known as Huygens Metasurfaces [15–18]. In this scenario, interference between electric and magnetic dipole scattered fields suppresses the backward radiation while enhancing the forward scattering. Combined with high non-resonant transmission through a materially sparse structure, the transmission efficiency can remain remarkably high at a fixed frequency while the transmission phase can span the full $2\pi$ range by varying some structural parameter. Another principle to realize a pure-phase resonance response in transmission is based on the concept of quasi-bound states in the continuum, and the associated polarization vortices, in thin-film photonic crystal structures [19–23]. In this case, certain parametric tuning associated with the vertical asymmetry of the structure slightly perturbs a bound state in the continuum so that it starts leaking radiation towards a preferred single radiation channel, hence causing an asymmetric scattering effect similar to the hybrid electric-magnetic dipole resonances if the preferred radiation channel is on the transmission side. Although these approaches are capable of producing almost ideal pure-phase resonances in transmission, maintaining very high transmission intensity while the phase is varied, they are fundamentally narrow-band effects and their parametric sensitivity is high in order to keep intricate coupling configurations optimally balanced. These features substantially restrict their practical applicability.

In this paper, we propose a new principle to realize a pure-phase resonance effect in transmission, which persists across a remarkably wide photonic band and is parametrically robust. Specifically, our approach is based on considering a glide-symmetric bilayer metasurface of which the synthetic band structure can be treated in analogy to a 2-dimensional (2D) Dirac semimetal [24,25]. We find that its characteristic eigenstates at a critical topological phase, band structure showing Dirac crossing, imply anomalous non-Hermitian orthogonality and extreme asymmetry in leakage-radiation distributions. These unique topological properties give rise to unidirectional resonance excitation and perfect leakage-radiation blazing at a single preferred radiation channel over nearly an entire photonic band, resulting in an extremely broadband pure-phase response in transmission. We provide a comprehensive theory and full-wave numerical analyses to elucidate this intriguing effect. Finally, to show the potential of these ideas in a relevant application scenario, we numerically demonstrate novel metasurfaces based on our proposed approach for efficient and broadband wavefront manipulation in transmission.

## II. GENERAL FEATURES OF PURE-PHASE RESONANCES

In a resonant scattering problem, the outgoing wave complex amplitude at an arbitrary radiation port can be expressed as

$$S = S_c + S_d(\omega) = S_c + A_d \frac{i\Gamma}{(\omega - \omega_0) + i\Gamma}, \quad (1)$$

Where $S_c$ is the amplitude of the non-resonant pathway through a broad continuum state, $S_d(\omega)$ is the amplitude of the resonant pathway through a narrow discrete state, $\omega$ is the frequency of the incident wave, $\omega_0$ is the center frequency of the discrete state excitation, and $\Gamma$ is the dissipation rate of the discrete state. This expression provides a simplified physical picture (valid for a single narrow resonance) according to which the non-resonant amplitude $S_c$ forms the spectral background and its interference with the resonant amplitude $S_d$ produces characteristic Fano line shapes in the intensity and phase of $S$. Note that $S_d$ is expressed in terms of resonance-excitation strength coefficient $A_d$ and normalized Lorentzian function $i\Gamma(\omega-\omega_0+i\Gamma)^{-1}$.

An ideal pure-phase resonance appears when the coupling configuration of a system satisfies the condition

$$A_d = -2S_c, \quad (2)$$

so that Eq. (1) reduces to

$$S = -S_c e^{2i\alpha(\omega)}, \quad (3)$$

where $\alpha(\omega) = \tan^{-1}[(\omega-\omega_0)/\Gamma]$. Clearly, Eq. (3) implies a frequency-independent, constant intensity, $I = |S|^2 = |S_c|^2$, and an abrupt $2\pi$ phase shift following $\arg(S) = \pi + \arg(S_c) + 2\cdot\tan^{-1}[(\omega-\omega_0)/\Gamma]$ while $\omega$ varies around $\omega_0$ over a sufficiently wide spectral region with respect to the resonance linewidth $\Gamma$. In this respect, creating a pure-phase resonance is simply the problem of finding a resonance structure where $S_c$ (non-resonant) and $A_d$ (resonant) are appropriately balanced at a certain desired port to satisfy Eq. (2).

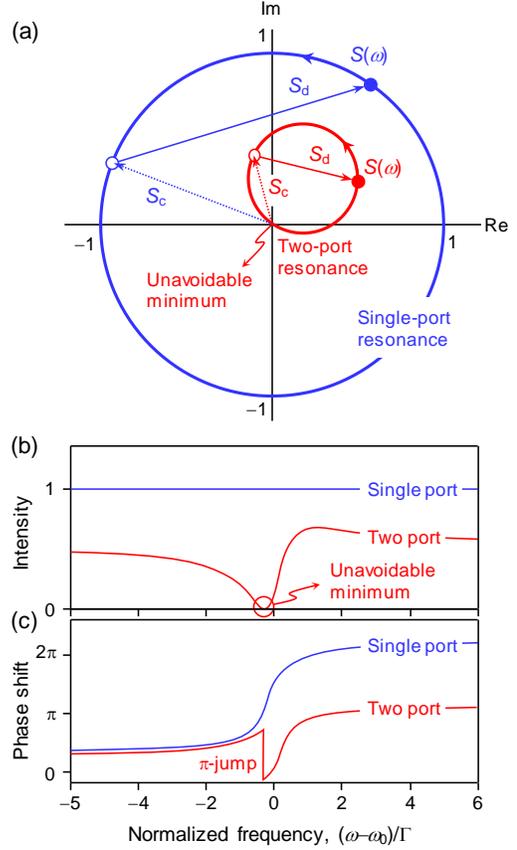

FIG. 1. General feature of pure-phase resonances. (a), Phasor diagrams of the typical $S(\omega)$ for a single radiation-port resonance and two radiation-port resonance. $S_c$ and $S_d$ are the non-resonant and resonant amplitudes, respectively, and $S$ is their linear superposition. (b), and (c), Typical intensity $|S|^2$ and phase $\arg(S)$ spectra for the single radiation-port and two radiation-port resonances. The two-port resonance involves an unavoidable intensity zero and an associated $\pi$-jump in the phase.

Such a condition can be easily obtained for a non-absorbing or largely over-coupled resonance state combined with a single port corresponding to a single input channel and a single output radiation channel (for example, a cavity with only a reflection channel, and no transmission, as in a GT etalon). In this case, temporal coupled-mode theory predicts $|\kappa|^2 = 2\Gamma$ and $A_d = -\kappa^2/\Gamma = -2\cdot\exp(2i\phi_\kappa)$ with $\phi_\kappa = \arg(\kappa)$, where $\kappa$ is the coupling constant between the port radiation and the resonance modes. Subsequently, the pure-phase resonance condition in Eq. (2) reduces to $S_c = \exp(2i\phi_\kappa)$. This implies that the ideal pure-phase response for a single radiation-channel resonance is obtained if $|S_c| = 1$ and $\arg(S_c) = 2i\phi_\kappa$. The strength condition of $|S_c| = 1$ is trivial for any non-absorbing system with a single output radiation channel due to energy conservation. Favorably, the phase condition of

arg($S_c$) = $2i\phi_\kappa$ is also naturally satisfied in the single port case because the time-reversal symmetry requirement of the scattering process dictates so in general [26,27].

Consider the blue circle in Fig. 1(a) for a phasor diagram of a typical $S(\omega)$ in a single-port resonance case. $S(\omega)$ is a superposition of the frequency-independent $S_c$ and the Lorentzian resonance amplitude $S_d(\omega)$, and it appears as a unit circle over which the phase shifts without any intensity change, as shown in Figs. 1(b) and 1C. This is exactly the case for the classical Gires-Tournois etalon [10] and for modern reflective dielectric metasurfaces [11–14] in which the zero-order reflection channel is the only output radiation channel for both the resonant and non-resonant pathways.

For a transmissive element supporting a *single* resonance, unfortunately, the single radiation-channel condition is not generally achievable because of the coexistence of reflection and transmission channels and the constraints imposed by the reciprocity and time-reversal symmetry of linear systems. The coexistence of two radiation channels is not simply a matter of imperfect intensity and phase distributions at some tolerable magnitudes. It greatly deteriorates applicability of the resonant phase modulation response due to an inevitable intensity zero, accompanying the π-jump in the phase. This intensity minimum is caused by a complete destructive interference between the non-resonant amplitude $S_c$ and the resonant amplitude $S_d$. As proved previously [26–27], the time-reversibility restriction for two output-port cases forces the phase difference between $S_d$ and $S_c$ to be π at a frequency point where $|S_d| = |S_c|$, making the destructive interference between them perfect at the intensity minimum (i.e., zero). A phasor diagram of the typical $S(\omega)$ for a two-port resonance is illustrated as the red circle in Fig. 1(a). It necessarily passes through the origin and the associated spectral profiles involve a zero in the intensity and a π-jump in the phase, seriously limiting its utility for phase modulation, as indicated in Figs. 1(b) and 1(c).

Considering these fundamental properties, an ideal transmissive pure-phase resonance is obtained only if the reflection channel is closed in both the non-resonant and resonant pathways. Previously, resonant reflection channel suppression has been obtained by using doubly resonant structures radiating in the same channels, e.g., arrays of nanoparticle meta-atoms where electric and magnetic dipole resonances are simultaneously excited [15–18]. In such structures, the backward scattering channel (reflection) can be suppressed by destructive interference between the electric and magnetic dipole radiation. In this way, one can design meta-atoms that cover the full 2π phase range in transmission, as a function of some structural parameters, while maintaining transmission intensity close to unity. However, this only works within a narrow spectral band in which the induced electric and magnetic dipole moment strengths and phases are coincidentally optimal. In addition, in order to produce an *ideal* transmissive pure-phase resonance, additional structural optimization and tuning should be performed to suppress also the non-resonant reflection channel without unfavorably affecting the resonant reflection suppression condition. Therefore, the doubly resonant nanoparticle array approach is fundamentally narrowband and demands careful parametric tuning of the unit cell structures and array geometries to obtain high performance and a robust pure-phase resonance response in transmission.

### III. EXTREMELY BROAD PURE-PHASE RESONANCE BAND

As a new platform to demonstrate a broadband pure-phase resonance response in transmission, we consider guided-mode resonances in a glide-symmetric bilayer grating structure, as shown in Fig. 2(a). Two identical grating layers are shifted from each other by a distance $a\Delta$, where $a$ is the period and $\Delta$ is the normalized glide shift. As conventional resonant gratings do, this structure produces narrow scattering resonances along dispersion bands at which any one of the diffraction orders has an in-plane phase distribution matching that of a leaky guided mode [25–31]. In the zero-order regime below the first-order Rayleigh frequency, a leaky guided mode couples to two radiation ports corresponding to the zero-order reflection and transmission plane-wave channels, and thereby gives rise to a two-port Fano-resonance profile with a typical response as indicated in Figs. 1(b) and 1(c).

Nevertheless, guided-mode resonances in the proposed structure, for certain optimal parameter values, produce an almost ideal pure-phase resonance band similar to what one would expect for single-port resonances, as shown in Figs. 2(b) and 2(c). See figure caption for the optimal parameter set that we considered for these results. In Fig. 2b, we show the transmittance ($T$) and transmission phase ($\phi_t$) spectra in the (in-plane) wavevector-frequency plane. In the lower vicinity of the first-order Rayleigh frequency $\nu_R(+1)$, we find a band of almost ideal pure-phase resonances, *i.e.*, constant intensity at $T \approx 1$ with full phase range $\Delta\phi_t \approx 2\pi$. Remarkably, this resonance persists over an extremely broad spectral band from $0.19c/a$ (108 THz) to $0.38c/a$ (217 THz). This band is 1,382-nm wide in wavelength and covers nearly the entire first-order guided-mode resonance band within the zero-order regime. Figure 2(c) shows phasor diagrams for the complex transmission coefficient $t(\lambda)$ at selected angles of incidence. For $\lambda_0 = 1.55$ μm, $t(\lambda)$ appears as a unit circle representing an ideal pure-phase resonance. Remarkably, the pure-phase resonance characteristic persists over a very broad spectral range including $\lambda_0 = 1.8$ and $2.15$ μm as indicated in the figure. If we define an operating wavelength range as the range supporting transmission resonances with the intensity minimum exceeding 50% while keeping a full 2π-phase shift, we obtain a 900-nm-wide range from 1.45 μm to 2.35 μm, *i.e.*, a 47% fractional wavelength bandwidth with respect to the average wavelength within that range. Importantly, the obtained operation range is an order of magnitude wider than the typical 3-5% fractional bandwidths obtained with previous approaches based on hybrid electric and magnetic dipole scattering (Huygens metasurfaces) [15,17] or polarization vortices associated with quasi-bound-state in the continuum [21].

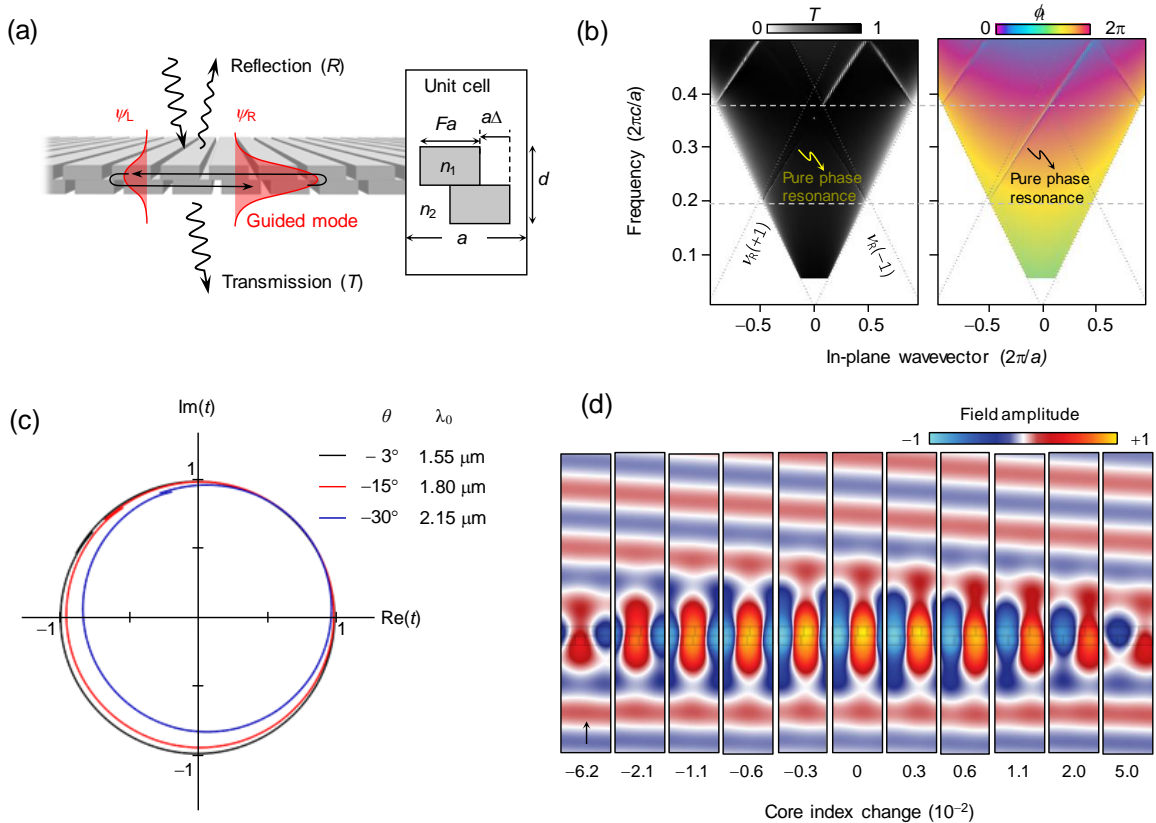

FIG. 2. Extremely broad pure-phase transmission resonance band in a glide-symmetric bilayer metasurface. (a), Schematic of the proposed structure. Two identical grating layers with period $a$, fill factor $F$, and core and cladding indices of refraction $n_1$ and $n_2$ are laterally shifted by a distance $a\Delta$. Guided modes propagating towards the left ($\psi_L$) and right ($\psi_R$) induce guided-mode resonances. (b), Transmittance ($T$) and transmission phase ($\phi$) spectra for an optimal structure under transverse-electric (TE) polarized light incidence. The parameters are $a = 525$ nm, $F = 0.8$, $d = 200$ nm, $n_1 = 3.48$ (Si), $n_2 = 2.45$ (Si$_3$N$_4$), and $\Delta = 1/4$. $\nu_R(\pm 1)$ indicates the first-order Rayleigh frequency. (c), Phasor diagrams of transmission coefficient $t(\lambda)$ at several selected angles of incidence, in which the range of $\lambda$ is ±100 nm with respect to the center wavelength $\lambda_0$. (d), Core-index-dependent transmitted wavefront change at the pure-phase resonance for $\lambda = 1.55$ μm and $\theta = -3°$.

This almost ideal pure-phase resonance in transmission is applicable for efficient wavefront control devices, e.g., metasurfaces, by suitably tuning its parameters. For example, we calculated transmitted field distributions as a function of core index change on the order of $10^{-2}$ and the results are shown in Fig. 2(d). A full one-wavelength shift in the transmitted wavefront is obtained without any significant change in the transmitted light intensity. Interestingly, this pure-phase resonance band appears to be accompanied by extreme asymmetry. Under TE illumination, it is produced only at the phase-matching condition for the guided mode ($\psi_R$) propagating towards the right ($+x$ direction), whereas no resonance feature is observed (see Fig. 2(b)) for the other guided mode ($\psi_L$) propagating towards the left ($-x$ direction).

The optimal parameter set to observe these effects is not obtained through blind optimization, but by requiring the system to satisfy the following conditions, whose physical significance is discussed in the following,

$$\text{Optimal film thickness } d \approx \pi j K_c^{-1}, \quad (4)$$

$$\text{Interlayer crossover coupling } \chi_1 = 0, \quad (5)$$

$$\text{Optimal glide shift } \Delta = \pm 1/4. \quad (6)$$

Here, $j$ is a positive integer and $K_c = (\varepsilon_c k_0^2 - k_x^2)^{1/2}$ is the effective out-of-plane wavevector component with $\varepsilon_c$ and $k_x$ denoting the effective dielectric constant of the grating layer and the in-plane wavevector of incident light, respectively. In addition, $\chi_1$ denotes the amplitude of coupling between the left- and right-propagating guided modes through two sequential first-order diffractions, one in the lower grating layer and another in the upper grating layer, thereby we refer it to as the interlayer crossover coupling. See Supplementary S1 for a detailed mathematical treatment leading to these conditions. If these conditions are satisfied, both the non-resonant and resonant reflection channels are simultaneously closed and the desired single radiation-channel condition is obtained. From a physical standpoint, Eq. (4) can simply be interpreted as an anti-reflection condition when the structure is treated as a homogeneous effective medium, thereby providing the necessary condition for the suppression of the non-resonant reflection-channel.

The necessary conditions for the suppression of the resonant reflection-channel, corresponding to Eqs. (5) and (6), are significantly more involved and are associated with unique interaction configurations at a glide-symmetric non-Hermitian topological nodal phase, as we will explain in detail in the next section.

## IV. NON-HERMITIAN TOPOLOGICAL PROPERTIES

In the spectral region of interest, guided modes propagating towards right ($\psi_R$) and left ($\psi_L$) are internally coupled to each other primarily through second-order diffraction processes and they are externally coupled to the radiation continuum primarily through first-order diffraction processes. Guided-mode eigenstates in the considered structure, at the vanishing interlayer crossover coupling condition in Eq. (5), can be described by a non-Hermitian binary eigenvalue problem

$$[\omega_c + i\gamma + (m+i\gamma)\cos(2\pi\Delta)\boldsymbol{\sigma}_1 + vk_x\boldsymbol{\sigma}_3]|\omega_\pm\rangle = \omega_\pm|\omega_\pm\rangle, \quad (7)$$

where $\boldsymbol{\sigma}_j$ is the Pauli matrix, $\omega_c$ is the center frequency of the second-order stop band, *i.e.*, the frequency of the second-order Bragg condition for the guided modes, $\gamma$ is the radiation rate of the guided modes in the absence of internal coupling, $m$ is the internal coupling constant, $v$ is the in-plane group speed of the guided modes, $k_x$ is the Bloch wavevector, and $|\omega_\pm\rangle = [\psi_{R\pm}\;\psi_{L\pm}]^T$. See Supplementary S1 for mathematical details of this description.

It is known from our recent work [24] that Eq. (7) allows us to treat the suggested glide-symmetric grating as a 2D Dirac semimetal which has its nodal point at $\Delta = \pm 1/4$. The nodal point condition exactly coincides with the conditions in Eqs. (5) and (6), strongly suggesting that the transmissive pure-phase resonance effect shares its underlying physics with 2D topological semimetals. Indeed, peculiar eigenstate properties and asymmetric leakage-radiation blazing effects provide a firm physical foundation for the almost ideal pure-phase resonances at the nodal-point condition.

First, the anomalous orthogonality of the eigenstates plays a key role. Eigenstates of non-Hermitian Hamiltonians are inevitably non-orthogonal in general scenarios. In our proposed structure, coupling of guided modes to the radiation continuum is an essential feature and the associated radiation loss results in non-Hermiticity, as parameterized by the leakage-radiation rate $\gamma$ of the guided mode in Eq. (7). Therefore, at the critical topological phase for $m = 0$ and $\Delta = 0$, the conventional Dirac point splits into two exceptional points and the eigenstates become non-orthogonal in a similar manner as in conventional topological, non-Hermitian, guided-mode resonance effects [32–34].

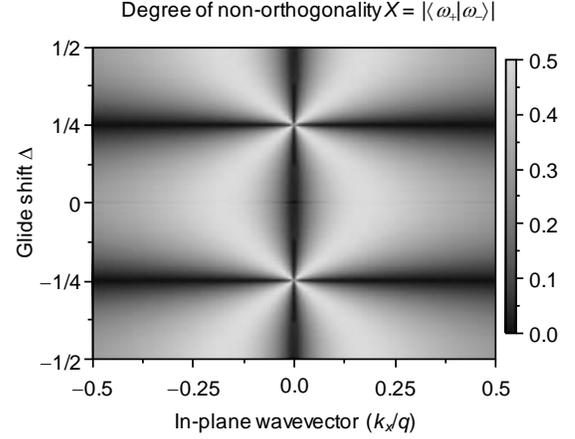

FIG. 3. Anomalous orthogonality of guided-mode eigenstates at the topological nodal phase. Degree of non-orthogonality as a measure of cross-projection magnitude of the two eigenstates are indicated as a function of glide-shift parameter $\Delta$ and in-plane wavevector $k_x$.

For $\Delta = \pm 1/4$, however, inter-guided-mode coupling represented by the term with $\boldsymbol{\sigma}_1$ in Eq. (7) vanishes. Consequently, the exceptional points merge into a Dirac point, the imaginary eigenvalue splitting vanishes to form a flat imaginary band, and the eigenstates become orthogonal throughout the entire frequency bands even in the presence of non-Hermiticity. We note that these atypical non-Hermitian properties are identical to those of 2D non-Hermitian topological semimetals [24,35].

We numerically confirmed the anomalous orthogonality in the considered case, as shown in Fig. 3(a). We indicate degree of non-orthogonality by taking the cross-projection magnitude $X = |\langle\omega_+|\omega_-\rangle|$ as a function of in-plane wavevector $k_x$ and glide shift parameter $\Delta$. $X$ vanishes throughout the entire $k_x$ domain at $\Delta = \pm 1/4$. This property is an immediate consequence of the emergence of the Dirac point.

In further detail, Eq. (7) implies an expression for the eigenstates as

$$|\omega_\pm\rangle = N(k_x)\begin{bmatrix}\sqrt{\Delta\omega \pm vk_x}\\ \pm\sqrt{\Delta\omega \mp vk_x}\end{bmatrix}, \quad (8)$$

where

$$\Delta\omega = \frac{1}{2}(\omega_+ - \omega_-) = \sqrt{(vk_x)^2 + (m+i\gamma)^2\cos^2(2\pi\Delta)}, \quad (9)$$

is the complex frequency splitting and $N(k_x) \approx |2\Delta\omega|^{-1/2}$ is a normalization constant. In the weak radiation-loss limit with $|m| \gg \gamma$, the degree of non-orthogonality is then expressed as

$$X = |\langle\omega_+|\omega_-\rangle| \simeq \frac{\gamma|vk_x\cos(2\pi\Delta)|}{(vk_x)^2 + m^2\cos^2(2\pi\Delta)}. \quad (10)$$

This expression conveniently explains the features observed in Fig. 3(a). The orthogonal eigenstate condition ($X$

= 0) is found at $k_x = 0$ (Γ-point) and $\Delta = \pm 1/4$. These conditions are evident in Eq. (10) because $X \propto |k_x \cdot \cos(2\pi\Delta)|$. The orthogonality at the Γ-point is also consistently understood by Eq. (8), which reduces to $|\omega_\pm\rangle = 2^{-1/2}[1\ \pm 1]^T$ at $k_x = 0$ regardless of $\Delta$ value.

At $\Delta = \pm 1/4$, the orthogonality persists over the entire Brillouin zone even in the presence of radiation loss ($\gamma \neq 0$) and, hence, we refer to this effect as "anomalous orthogonality". This is an immediate consequence of the decoupling between the right and left guided modes ($\psi_R$ and $\psi_L$) at the topological nodal-phase condition. At this special condition, $\Delta\omega = |vk_x|$ according to Eq. (9) and its application to the eigenstate expression in Eq. (8) results in

$$|\omega_\pm\rangle = \begin{bmatrix}1\\0\end{bmatrix} \text{ or } \begin{bmatrix}0\\1\end{bmatrix}. \quad (11)$$

The decoupled orthogonal eigenstates at this specific condition have no dependence on $k_x$ and, as discussed in the following, extremely asymmetric responses might be created over the entire resonance band. Here, it appears that the lattice structure at $\Delta = \pm 1/4$ does not provide any net effect on the guided modes because the inter-guided-mode coupling vanishes. However, this specific lattice structure has remarkable effects on the leakage radiation distribution, leading to an almost ideal pure-phase resonance as explained in the next section.

## V. PERFECT BLAZING IN LEAKAGE RADIATION

The decoupled guided-mode states and their interaction with the radiation continuum at $\Delta = \pm 1/4$ induce perfect blazing of leakage radiation towards a single preferred diffraction order and, consequently, lead to the effective realization of a single-port resonance in transmission.

In Figs. 4(a)~4(e), we show $\Delta$-dependent frequency bands and their partial leakage-radiation strengths towards the cover ($\kappa_c$) and substrate ($\kappa_s$) media for the upper ($\omega_+$) and lower ($\omega_-$) band states. For $\Delta = 0$ or $1/2$, $\omega_\pm$ bands have a typical anti-crossing structure and the partial leakage-radiation strengths are symmetric along both vertical and lateral axes, i.e., $|\kappa_c| = |\kappa_s|$ and $\kappa_{c\text{ or }s}(k_x) = \kappa_{c\text{ or }s}(-k_x)$. In the other cases for $0 < \Delta < 1/2$, $|\kappa_c| \neq |\kappa_s|$ and $\kappa_{c\text{ or }s}(k_x) \neq \kappa_{c\text{ or }s}(-k_x)$ in general. This up/down and left/right asymmetry in the leakage-radiation distribution can be intuitively understood as resulting from the broken mirror symmetry of the structure with respect to x-y and y-z planes.

Importantly, we see several very peculiar characteristics in these cases. First, the leakage-radiation distribution is extremely asymmetric at $\Delta = 1/4$. The frequency bands show a typical Dirac point appearing at the topological nodal phase resulting from complete decoupling between the right and left guided modes, i.e., $|\omega_R\rangle = |R\rangle = [1\ 0]^T$ and $|\omega_L\rangle = |L\rangle = [0\ 1]^T$ over the entire frequency domain. The leakage-radiation distributions of these completely decoupled guided-mode eigenstates are also completely decoupled in such a way that $\kappa_c = 0$ (no radiation towards the cover) for $|\omega_R\rangle$ while $\kappa_s = 0$ (no radiation towards the substrate) for $|\omega_L\rangle$. We refer to this effect as asymmetric perfect leakage-radiation blazing in a similar sense to the classical blazed gratings that maximize the diffraction efficiency at a selected diffraction order by matching ray-optical and diffractive-optical angles. In our case, however, the blazing happens for the cover and substrate zero-order waves, not different diffraction orders, and its underlying physics is different from the classical blazed gratings.

Although the perfect leakage-radiation blazing appears throughout the entire frequency bands for $\Delta = 1/4$, strong blazing effect is still persistent even for $\Delta \neq 1/4$, as shown in Figs. 4(b) and 4(d). In particular, for these intermediate conditions, almost perfect blazing is found at two opposite $k_x$ points on the $\omega_-$ band. Specifically, looking at the radiation rate spectra for the $\omega_-$ band in Figs. 4(b) and 4(d), we see $\kappa_s \approx 0$ or $\kappa_c \approx 0$ at $k_x = \pm 0.01 \times 2\pi/a$. The almost perfect blazing for $\Delta \neq 1/4$ at these points is associated with the presence of a bound state in the continuum (BIC) located at the Γ-point ($k_x = 0$) for $\Delta = 0$ and $1/2$, as further elucidated below.

In order to better understand the leakage-radiation blazing conditions, we calculate ratio of the leakage-radiation strengths $|\kappa_c/\kappa_s|$ for $\omega_+$ and $\omega_-$ bands on the $k_x$-$\Delta$ plane as shown in Figs. 4(f) and 4(g), respectively. The perfect blazing effect over the entire bands is clearly seen at $\Delta = 1/4$, where the ratio $|\kappa_c/\kappa_s|$ is maximized (radiation toward cover) or minimized (radiation toward substrate). In Fig. 4(h), we show excited eigenstates and their leakage radiation distributions at $\Delta = 1/4$ at four selected conditions I~IV, indicated in Figs. 4(c), 4(d) and 4(g). We confirm that the perfect blazing is closely related to the eigenstates $|R\rangle$ and $|L\rangle$ such that $|L\rangle$ emits radiation only towards the cover while $|R\rangle$ emits radiation only towards the substrate. For the $\omega_-$ band in Fig. 4(g), the map areas for $\Delta \neq 1/4$ reveal additional strong blazing conditions ($|\kappa_c/\kappa_s| > 10^{+2}$ or $|\kappa_c/\kappa_s| < 10^{-2}$) which are connected to the BIC at $\Delta = 0$ and $\pm 1/2$.

To further elucidate this behavior, we can use the coupled-mode theory by Kazarinov and Henry [28] to determine closed-form expressions for the leakage radiation strength as

$$|\kappa_{c\pm}| = \sqrt{2\gamma}\,|\psi_{L\pm}\cos[\pi(\Delta-1/4)] - \psi_{R\pm}\sin[\pi(\Delta-1/4)]|, \quad (12)$$

$$|\kappa_{s\pm}| = \sqrt{2\gamma}\,|\psi_{R\pm}\cos[\pi(\Delta-1/4)] - \psi_{L\pm}\sin[\pi(\Delta-1/4)]|, \quad (13)$$

and these expressions provide comprehensive understanding of the leakage-radiation blazing properties revealed in the numerical analyses. Here, $\psi_{R\pm} = \langle R|\omega_\pm\rangle$ and $\psi_{L\pm} = \langle L|\omega_\pm\rangle$ and the expressions are derived by taking spatial symmetries and interferences between the two grating layers into account. See Supplementary S2 for detailed mathematical treatment. Note in Eqs. (12) and (13) that the spatial-symmetry property $\psi_{L\pm}(-k_x) = \psi_{R\pm}(+k_x)$ of the eigenstates results in $|\kappa_{c\pm}(-k_x)| =$ $|\kappa_{s\pm}(+k_x)|$. Therefore, the blazing conditions for the cover and substrate leakage radiations symmetrically appear with respect to the Γ-point, as found in Figs. 4(f) and 4(g).

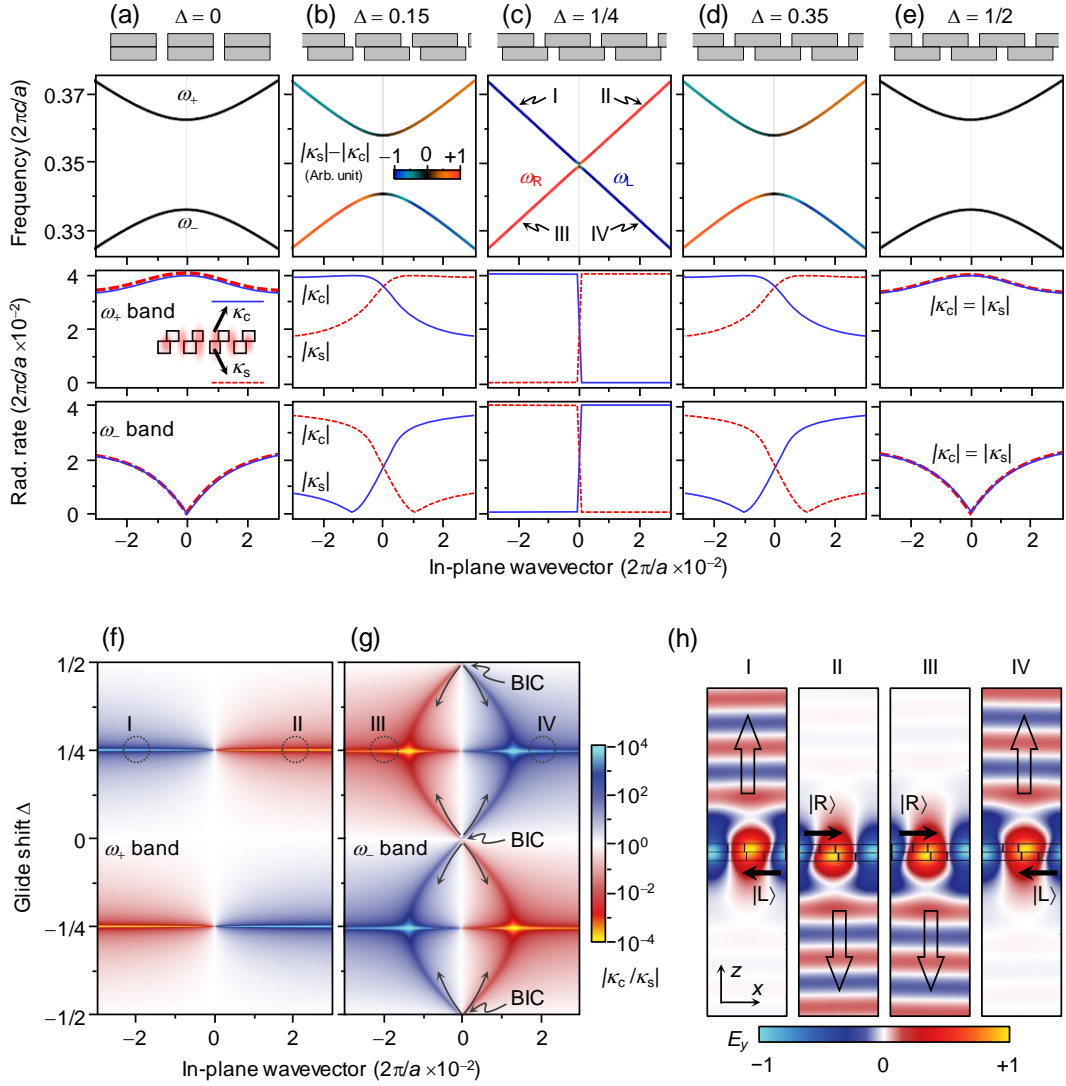

FIG. 4. Perfect blazing of leakage-radiation. (a)~(e), Glide-shift-dependent photonic bands (top) and corresponding leakage-radiation rates towards the cover (middle) and substrate (bottom) near the second-order Bragg condition (stop band). $\omega_\pm$ denotes the upper (+) and lower (−) band, respectively. $\kappa_c$ and $\kappa_s$ represent leakage-radiation rates towards the cover and substrate, respectively. (f) and (g), Leakage-radiation asymmetry ratio $|\kappa_c/\kappa_s|$ as a function of glide shift parameter $\Delta$ and in-plane wavevector $k_x$ for $\omega_+$ (f) and $\omega_-$ (g) bands. (h), Field distributions for four selected perfect leakage-radiation blazing conditions I, II, III, and IV. Parametric locations of these conditions are indicated in panels (c), (f), and (g).

For $\Delta = \pm 1/4$, Eqs. (12) and (13) reduces to

$$\begin{cases} |\kappa_{c\pm}| = \sqrt{2\gamma}|\psi_{L\pm}| \ \& \ |\kappa_{s\pm}| = \sqrt{2\gamma}|\psi_{R\pm}| \text{ for } \Delta = 1/4 \\ |\kappa_{c\pm}| = \sqrt{2\gamma}|\psi_{R\pm}| \ \& \ |\kappa_{s\pm}| = \sqrt{2\gamma}|\psi_{L\pm}| \text{ for } \Delta = -1/4 \end{cases}. \quad (14)$$

Clearly, for $\Delta = 1/4$, the leakage-radiation towards the cover is exclusively from $|L\rangle$ while the leakage-radiation towards the substrate is solely from $|R\rangle$, as seen in Fig. 4(h). Therefore, perfect blazing at the cover or substrate zero-order occurs for resonance band $\omega_R$ or $\omega_L$ that takes $|R\rangle$ or $|L\rangle$ as the eigenstate, respectively.

For $\Delta \neq \pm 1/4$ in contrast, Eqs. (12) and (13) imply that perfect blazing ($\kappa_c = 0$ or $\kappa_s = 0$) is obtained for the following conditions

$$\frac{\psi_{L\pm}}{\psi_{R\pm}} = \pm\sqrt{\frac{\Delta\omega \pm vk_x}{\Delta\omega \mp vk_x}} = \begin{cases} \tan[\pi(\Delta - 1/4)] \text{ for } \kappa_{c\pm} = 0 \\ \cot[\pi(\Delta - 1/4)] \text{ for } \kappa_{s\pm} = 0 \end{cases}. \quad (15)$$

Eq. (15) quantitatively describes the strong blazing conditions connected to the BICs in Fig. 4(g). In the close vicinity of the BIC that is located at the band edge of the $\omega_-$ band for $\Delta \approx 0$, for example, the blazing condition of Eq. (15) reduces to $vk_x/\Delta\omega \approx 2\pi\Delta$ for $\kappa_{c-} = 0$ or $vk_x/\Delta\omega \approx -2\pi\Delta$ for $\kappa_{s-} = 0$. This is exactly consistent with the feature revealed at the bottom of Fig. 4(g) where the blazing conditions for the cover and substrate leakage-radiations linearly branch off from the $\Gamma$ point towards the left and right, respectively, as $\Delta$ increases from 0. In addition, one can easily find that there is no sensible solution to Eq. (15) for the $\omega_+$ band, which explains why the blazing condition for the $\omega_+$ band appears only at $\Delta = \pm 1/4$, as shown in Fig. 4(f).

These leakage-radiation properties described by Eqs. (12) and (13) enable a simple and intuitive model that quantitatively describes the transmission intensity and phase spectra. Specifically, we construct a temporal coupled-mode model of scattering amplitudes. For the non-resonant scattering processes in this model, we treat the grating structure as an effective homogeneous thin film. For the resonant scattering processes on the other hand, we take the binary eigenstates in Eq. (7) as resonance states. We assume that coupling strengths between the resonance states and radiation channels are given by Eqs. (12) and (13). Although the model only assumes time-reversal symmetry, reciprocity, and energy conservation for the states given by Eq. (7), with coupling strengths in Eqs. (12) and (13), the calculation results for the angle-dependent intensity and phase spectra in transmission and reflection are quantitatively consistent with rigorous numerical analysis based on the finite-element method. See Supplementary S3 for a detailed mathematical derivation and its comparison with numerical calculation results.

In particular, the model predicts transmission coefficients for cover-side light incidence

$$\tau_{14} = it + \frac{\kappa_1^2}{i(\omega-\omega_0)-\Gamma}, \quad (16)$$

$$\tau_{23} = it + \frac{\kappa_2^2}{i(\omega-\omega_0)-\Gamma}, \quad (17)$$

where $\tau_{14}$ and $\tau_{23}$ are transmission coefficients for positive $k_x$ and $-k_x$, respectively. Here, $it$ represents the non-resonant transmission coefficient, $\kappa_1$ and $\kappa_2$ are coupling constants of the resonance states with the corresponding transmission channels, $\omega_0 = \text{Re}(\omega_\pm)$ is the resonance center frequency, and $\Gamma = \text{Im}(\omega_\pm)$ is the decay rate of the resonance state. For the optimal conditions of $\Delta = 1/4$ and Eq. (5), the parameters take values as $t \approx 1$, $\kappa_1 \approx (2\gamma)^{1/2}\exp(i\pi/4)$, $\kappa_2 \approx 0$, and $\Gamma \approx \gamma$. See Supplementary S3 for the associated closed-form expressions of these parameters. Then, $\tau_{14}$ in Eq. (16) for the transmission towards the perfectly blazed channel reduces to

$$\tau_{14} \approx i + \frac{2\gamma}{\omega-\omega_0+i\gamma} = i\frac{\omega-\omega_0-i\gamma}{\omega-\omega_0+i\gamma}. \quad (18)$$

This is identical to Eq. (1) with non-resonant amplitude $S_c = i$ and $A_d = -2i$, and hence it satisfies the ideal pure-phase resonance condition in Eq. (2), i.e., $A_d = -2S_c$. Note that the right-hand side of Eq. (18) clearly represents frequency-independent constant intensity at $|\tau_{14}| = 1$ and frequency-dependent phase $\arg(\tau_{14}) = \pi/2 - \tan^{-1}[(\omega-\omega_0)/\gamma]$, which is expected for a phasor along the unit circle as indicated by the blue circle in Fig. 1(a) for a single-port resonant-scattering amplitude. Thus, rather strikingly, the considered transmissive element behaves exactly as an open resonator connected to a single radiation channel over a very broad bandwidth. In contrast, $\tau_{23}$ in Eq. (17) for the transmission towards the closed channel reduces to $\tau_{23} \approx i = constant$ which does not have any frequency-dependent resonance feature in the intensity and phase. Therefore, our topological-state and leakage-radiation theory of the bilayer glide-symmetric grating structure fully captures the underlying physics of the asymmetric pure-phase resonance effect in transmission.

## VI. POTENTIAL APPLICATION

Combined with appropriate index-tuning mechanisms or spatial parameter distributions, i.e., suitable spatial structuring, the pure-phase transmission resonance may have a wide range of application areas as a compact flat optical device, such as phase modulators and wavefront control metasurfaces. Considering the wavefront engineering capability in particular, precise manipulation of local phase delay is a key feature. Although the proposed pure-phase resonance effect in transmission provides an efficient mean for controlling global phase delay over the cross-section of an optical beam in the absence of undesirable intensity variations, it is unclear whether the phase delay in the proposed structure is locally controllable within an acceptable level in practice. This question naturally arises because the guided-mode resonance is basically a nonlocal [36] and collective response involving multiple unit cells typically in the order of 10 to 1000.

In this respect, we performed a set of numerical experiments on metasurface lensing functionality based on the proposed pure-phase resonance effect. We apply a quadratic phase delay distribution (which approximates well an exact hyperbolic phase distribution for small distances from the center) as

$$\phi(x) = \phi_0 - \frac{\pi x^2}{\lambda f}, \quad (19)$$

where $\phi_0$ is a reference phase delay at the center ($x = 0$) and $f$ is the desired focal length. This quadratic phase-delay distribution is wrapped within a $[0, 2\pi)$ range, resulting in piece-wise continuous Fresnel zones. The wrapped phase-delay distribution is coded into a corresponding period distribution $a(x; f)$.

We first tested a trial design based on the structure considered in Fig. 1. We assume a 80-μm-wide lens for $f = 40$, 60, and 80 μm at wavelength 1.5 μm. Unfortunately, we found no clear signature of a focal spot in the FEM simulation results. We attribute the failure in these cases to the strong nonlocal response of the structure, as expected. Effective propagation length of the guided mode in these first trial cases is $L_{\mathrm{eff}} \sim 230a$, which we estimate from the resonance quality factor $Q \sim 4{,}400$ and group speed $v \sim 0.35c$ using the relation $L_{\mathrm{eff}} \approx (2\pi)^{-1} \cdot Q \cdot (v/c) \cdot a$, which is approximately valid for resonant gratings around the first-order Bragg condition in general. It is therefore not surprising to see no local phase-delay effect in this high-$Q$ and long-propagation-length regime.

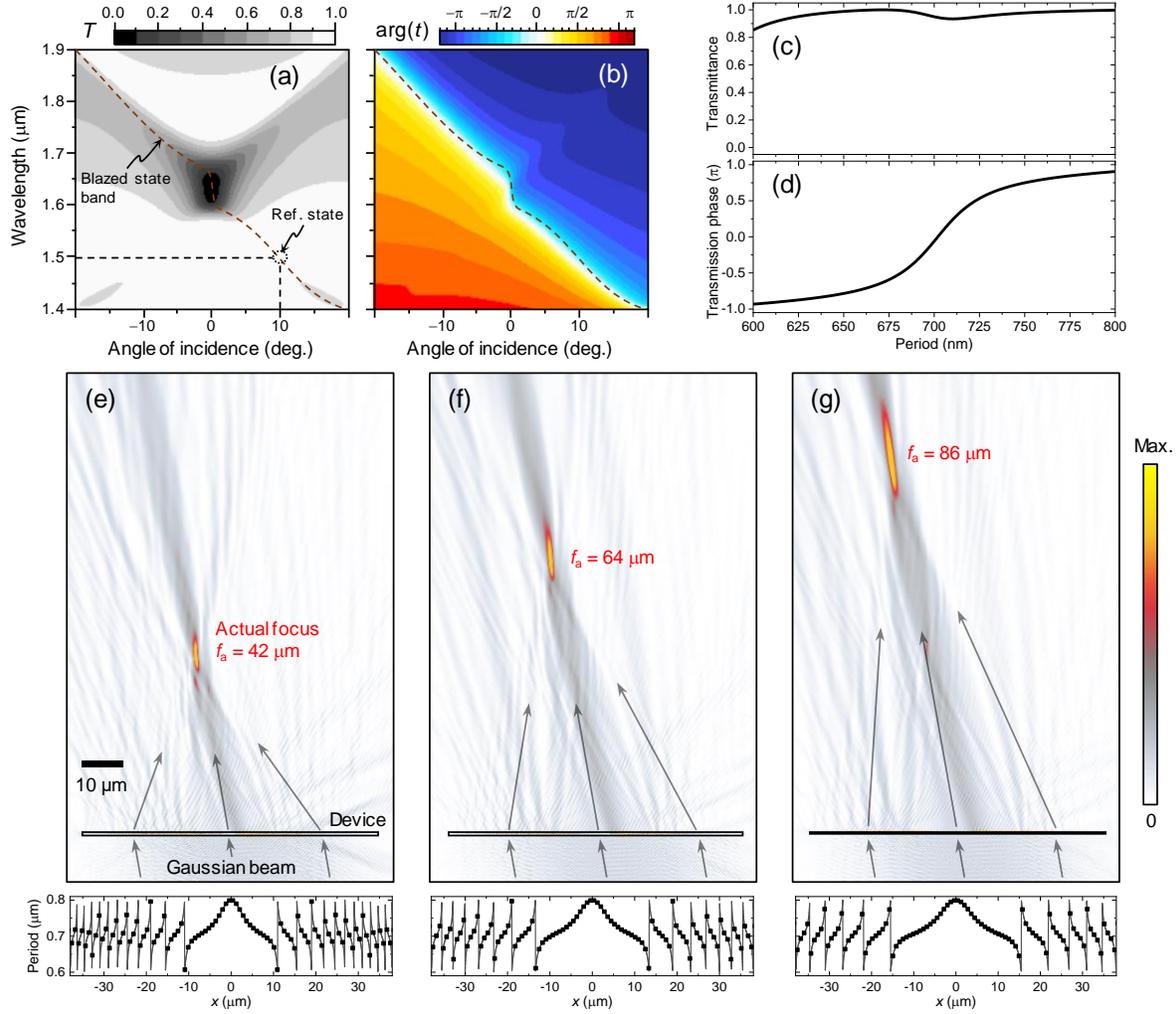

FIG. 5. Numerical demonstration wavefront control. (a) and (b), Transmittance $T$ and transmission phase $\arg(t)$ spectra for a structure with parameters $a = 700$ nm, $F = 0.6$, $d = 250$ nm, $n_1 = 3.48$ (Si), and $n_2 = 1.5$ (SiO$_2$). (c) and (d), $T$ and $\arg(t)$ as a function of period for a pure-phase resonance condition at wavelength $\lambda = 1.5$ μm and angle of incidence $\theta = 10°$. (e)~(g), Focusing of an incident Gaussian beam for desired focal length $f_d$ = 40, 60, and 80 μm. The period distribution on the bottom of each panel is configured so that the device yields the desired focal length in the ideal local-response regime. $f_a$ denotes focal length obtained through simulation.

In order to significantly reduce this effective propagation length to enable a sufficiently stronger local phase-delay response, we modify the structure. We change the host medium from Si$_3$N$_4$ ($n_2 = 2.45$) to SiO$_2$ ($n_2 = 1.5$) to increase the index contrast. In addition, we modify the thickness of the Si grating layer from 200 nm to 250 nm and fill factor from 0.8 to 0.6. These modifications are favorable for increasing the radiation decay rate of the guided mode since the confinement factor of the guided mode to the grating layer and the diffraction strength are enhanced with such parametric changes. With these modified design parameters, the resonance quality factor and group speed for this modified structure are $Q = 43$ and $v = 0.37c$, resulting in effective propagation length $L_{\mathrm{eff}} \sim 2.5a$, which seems very promising to expect an improved local phase-delay response in contrast to the original structure with $L_{\mathrm{eff}} \sim 230a$.

In Figs. 5(a) and 5(b), we show transmission intensity $T$ and phase delay $\arg(t)$ spectra. Along the blazed-state resonance condition indicated by the dark-red dashed curve, the $T$ spectrum is persistently high above 0.8 over most of the spectral domain of interest, except for the $\Gamma$ point, while maintaining a $2\pi$ phase shift around it, as expected. For a reference resonance state at 1.5 μm wavelength and 10° angle of incidence, we calculate $T$ and arg($t$) as the grating period $a$ changes from 600 to 800 nm, as shown in Figs. 5(c) and 5(d). These results confirm $T > 0.8$ over the entire period range while arg($t$) changes over almost a full $2\pi$ span. Using this period-dependent response, we then create a metasurface with a period distribution that realizes a phase profile according to Eq. (19) for the desired focal length $f$ = 40, 60, and 80 μm. The simulation results are provided in Figs. 5(e) ~ 5(g). We show the designed period distributions on the bottom and the corresponding transmitted field intensity distributions on top. The incident Gaussian beam is 40-μm-wide in its full-width at half-maximum for all three cases. As seen clearly, we obtain well-defined focal spots for all cases. The focal-length errors are around 6% from their desired values. We consider that this error is related to the nonlocal response of the resonant transmission effect, as we discussed earlier in this section. Besides the focal-length error, these simulation results suggest highly promising performance as the reflection of the incident Gaussian beam is lower than 1.7% and the power efficiency of the focal spot, *i.e.*, portion of the optical power contained within the focal spot area, is higher than 60%. Moreover, different from Huygens metasurfaces based on localized magnetic-electric dipole resonances, our design guarantees similar performance metrics over a very broad wavelength range, albeit for different angles of incidence at different wavelengths, as suggested by the results in Figs. 5(a), (b).

## VII. CONCLUSION

In conclusion, we have proposed a Dirac bilayer metasurface in a topological nodal phase as a novel, unique platform to enable pure-phase resonances in transmission with their operating band persisting over an extremely wide spectral window, as opposed to previously proposed structures. We have developed a complete theory to study the topological guided-mode eigenstates, the glide-symmetric leakage-radiation properties, and their contribution in the resonant scattering process. Based on this theoretical framework, we have found that the transmissive pure-phase resonance is enabled by the anomalous non-Hermitian orthogonality of the eigenstates and a perfect leakage-radiation blazing effect, which are, in principle, inherently broadband. We confirm quantitative agreement between our theory and rigorous numerical calculations. Finally, we numerically demonstrate a wavefront-controlling metasurface functioning as a transmissive microlens with remarkably high efficiency.

From a fundamental viewpoint, the proposed structure is asymmetric along both the vertical and lateral axes and thereby an asymmetric response may be expected. Nevertheless, our theory suggests that appropriate choices of a structural parameter set, within moderate value ranges, can induce extremely asymmetric responses at certain critical conditions associated with the topological phase of a structured medium. This implies that the proposed effect should be robust against imperfections and parametric drifts generally unavoidable in practice. Therefore, the proposed structure and its underlying physics should be applicable for the realization of various, highly robust, transmissive metasurface elements. Immediate application areas are phase modulators and wavefront control devices such as electro-optic phase shifters, solid-state beam steering elements, orbital-angular-momentum generators, and various holographic optical elements, which can take technical advantages of operating efficiently in transmission mode. In addition, the anomalous non-Hermitian orthogonality and leakage-radiation blazing effects may find their own independent applications including unidirectional or anti-reflecting couplers with virtually maximal efficiencies.

From another perspective, the extremely asymmetric responses in our system share their underlying physics with the 2D Dirac semimetal phase. This suggests that similar or related effects are potentially obtained in other topological systems such as Weyl-semimetal-like systems as 3-dimensional (3D) extension of the 2D Dirac semimetal [37,38]. We believe it is of great interest to further investigate theoretically and experimentally how the extreme asymmetry in such a 2D topological system manifests in its 3D extensions and how it may open new opportunities for applications.

## ACKNOWLEDGEMENTS

This research was supported in part by the Leader Researcher Program (NRF-2019R1A3B2068083), and the research fund of Hanyang University (HY-202000000000513). K.Y.L. acknowledges the NRF Sejong Science fellowship (NRF-2022R1C1C2006290) funded by the MSIT of the Korean government. F.M acknowledges support from the U.S. Air Force Office of Scientific Research (FA9550-22-1-0204).

# Supplementary Materials for

## Dirac Bilayer Metasurfaces as an Inverse Gires-Tournois Etalon

Ki Young Lee[1,2], Kwang Wook Yoo[1], Francesco Monticone[3], and Jae Woong Yoon[1,2],*

*Corresponding author. Email: yoonjw@hanyang.ac.kr

**This PDF file includes:**

Supplementary Text

Figs. S1 to S6

## S1. Eigenvalue problem for guided-mode resonances in a glide-symmetric bilayer grating

We assume a transverse-electric (TE) planar mount problem where a waveguide grating lies in x-y plane with its axis of periodicity in x axis and the plane of incidence is z-x plane. The transversal component of the electric field $E_y$ is described by the two-dimensional Helmholtz equation

$$\left[\frac{\partial^2}{\partial x^2} + \frac{\partial^2}{\partial z^2} + \varepsilon(x,z)\frac{\omega^2}{c^2}\right]E_y(x,z) = 0 \tag{S1}$$

with dielectric function

$$\varepsilon(x,z) = \begin{cases} \varepsilon_d & z > +h \\ \sum_{n=-\infty}^{\infty} \varepsilon_n(z)e^{inqx} & -h \leq z \leq +h \\ \varepsilon_d & z < -h \end{cases} \tag{S2}$$

$$\varepsilon_n(z) = \Delta\varepsilon_n \times \begin{cases} e^{+i\pi n\Delta} & (0 < z \leq +h) \\ e^{-i\pi n\Delta} & (-h < z \leq 0) \end{cases} \tag{S3}$$

$$\Delta\varepsilon_n = F(\varepsilon_H - \varepsilon_L)\text{sinc}(nF) + \varepsilon_L \delta_{0n} \tag{S4}$$

where we use normalized sinc function $\text{sinc}(X) = \sin(\pi X)/(\pi X)$.

Solving Eq. (S1), we take an ansatz as

$$E_y(x,z) = \left[L(z) + \psi_L(k_x)u(z)e^{-iqx} + \psi_R(k_x)u(z)e^{+iqx}\right]\exp ik_x x, \tag{S5}$$

where transversal wave function $u(z)$ of the fundamental guided mode is determined by

$$\left[\frac{d^2}{dz^2} + \varepsilon_0(z)\frac{\omega_u^2}{c^2} - \beta^2\right]u(z) = 0. \tag{S6}$$

This ansatz describes the electromagnetic responses in terms of interactions between zero-order wave $L(z)$, guided mode $\psi_L(k_x)u(z)e^{-iqx}$ propagating towards –x (left) direction, and guided mode $\psi_R(k_x)u(z)e^{-iqx}$ propagating towards +x (right) direction for given Bloch wavevector $k_x$. Here, zero-order wave $L(z)$ involves incident wave from a certain external plane wave source and leakage radiation from the guided modes in general. Following the coupled-mode formalism due to Kazarinov and Henry for these structural and solution settings in the vicinity of the second-order Bragg condition of $q = \beta$, we obtain an eigenvalue problem

$$\begin{bmatrix} \omega_c + X_0 + vk_x & X_2 - Z_2 \\ X_2 - Z_2 & \omega_c + X_0 - vk_x \end{bmatrix}\begin{bmatrix} \psi_R \\ \psi_L \end{bmatrix} = \omega(k_x)\begin{bmatrix} \psi_R \\ \psi_L \end{bmatrix} \tag{S7}$$

with $v$ denoting group speed of the guided mode and $\omega_c$ being the second-order Bragg condition frequency where $\beta = q$. The parameters in the matrix on the right-hand side are determined by following relations

$$X_0 = \frac{\omega_c}{2}\left[\zeta_0^{-1}k_c^2\int_{-h}^{+h}dz\int_{-h}^{+h}dz'\,u(z)^*\varepsilon_{\pm 1}(z)G(z,z')\Delta\varepsilon_{\mp 1}(z')u(z')\right], \tag{S8}$$

$$X_2 = \frac{\omega_c}{2}\left[\zeta_0^{-1}k_c^2\int_{-h}^{+h}dz\int_{-h}^{+h}dz'\,u(z)^*\varepsilon_{\pm 1}(z)G(z,z')\Delta\varepsilon_{\pm 1}(z')u(z')\right], \tag{S9}$$

$$Z_2 = \frac{\omega_c}{2}\left[\zeta_0^{-1}\int_{-h}^{+h}dz\,u(z)^*\varepsilon_{\pm 2}(z)u(z)\right], \tag{S10}$$

$$\zeta_0 = \int_{-\infty}^{+\infty} dz\, u(z)^* \varepsilon_0(z) u(z), \tag{S11}$$

$$G(z,z') \atop (-h \le z' \le h) = \frac{1}{2iK_c} \frac{1}{1-\rho^2} \times \begin{cases} \tau e^{iK_d h} e^{+iK_d z} & e^{-iK_c z'} + \rho e^{+iK_c z'} & (z > h) \\ e^{+iK_c z} + \rho e^{-iK_c z} & e^{-iK_c z'} + \rho e^{+iK_c z'} & (-h \le z' < z \le h) \\ e^{-iK_c z} + \rho e^{+iK_c z} & e^{+iK_c z'} + \rho e^{-iK_c z'} & (-h \le z < z' \le h) \\ \tau e^{iK_d h} e^{-iK_d z} & e^{+iK_c z'} + \rho e^{-iK_c z'} & (z < -h) \end{cases}, \tag{S12}$$

where $\rho = r\exp(2iK_c h)$, $\tau = t\exp(iK_c h)$, $K_c = (\varepsilon_c k_0^2 - k_x^2)^{1/2}$ and $K_d = (\varepsilon_d k_0^2 - k_x^2)^{1/2}$, $k_0 = \omega/c$, $\varepsilon_c = \Delta\varepsilon_0 + \varepsilon_L$, and $r$ and $t$ are the internal reflection and transmission coefficients at the boundary surfaces of the waveguide layer. Note that Green's function $G(z, z')$ is a solution of an one-dimensional impulse response wave equation

$$\left[\frac{d^2}{dz^2} + \varepsilon_0(z)k_0^2 - k_x^2\right]G(z,z') = \delta(z-z'). \tag{S13}$$

Further considering these parameters in relation with the given expressions of the glide-symmetric dielectric function and the specific vertical symmetries of the associated functions such as $G(-z, -z') = G(z', z)$ and $u(-z) = u(z)$, the eigenvalue equation in Eq. (S7) can be rewritten by an alternative form explicitly revealing its dependence on normalized grating shift $\Delta$ as

$$\begin{bmatrix} \omega_c + \alpha_1 + \chi_1 \cos(2\pi\Delta) + vk_x & (\alpha_1 - \alpha_2)\cos(2\pi\Delta) + \chi_1 \\ (\alpha_1 - \alpha_2)\cos(2\pi\Delta) + \chi_1 & \omega_c + \alpha_1 + \chi_1 \cos(2\pi\Delta) - vk_x \end{bmatrix} \begin{bmatrix} \psi_R \\ \psi_L \end{bmatrix} = \omega(k_x) \begin{bmatrix} \psi_R \\ \psi_L \end{bmatrix}, \tag{S14}$$

where

$$\alpha_1 = \omega_c\, \Delta\varepsilon_1^2 \times \frac{k_c^2}{2\zeta_0} \begin{bmatrix} \int_0^h dz \int_0^h dz'\, u(z)^* G(z,z') u(z') \\ + \int_{-h}^0 dz \int_{-h}^0 dz'\, u(z)^* G(z,z') u(z') \end{bmatrix} \tag{S15}$$

$$\chi_1 = \omega_c\, \Delta\varepsilon_1^2 \times \frac{k_c^2}{2\zeta_0} \begin{bmatrix} \int_0^h dz \int_{-h}^0 dz'\, u(z)^* G(z,z') u(z') \\ + \int_{-h}^0 dz \int_0^h dz'\, u(z)^* G(z,z') u(z') \end{bmatrix} \tag{S16}$$

$$\alpha_2 = \omega_c\, \Delta\varepsilon_2 \times \frac{1}{2\zeta_0} \left[\int_{-h}^h dz\, u(z)^* u(z)\right] \tag{S17}$$

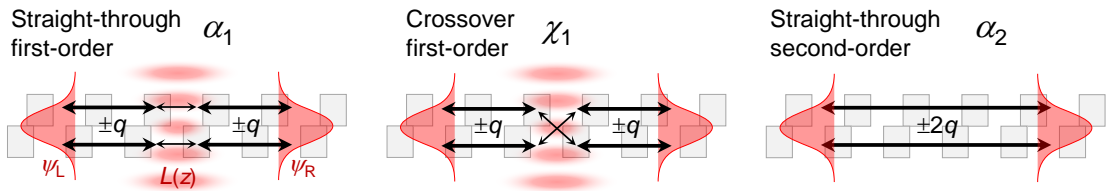

**Figure S1.** Three distinguished pathways of inter-modal coupling constants in Eqs. (S15) ~ (S17).

Expressions on the right-hand sides of Eqs. (S15)~(S17) allows intuitive physical interpretations of these parameters as following descriptions.

- *Straight-through first-order coupling rate* $\alpha_1$: Inter-modal coupling through the two consecutive first-order diffraction processes happening in each of the two grating layers.

- *Crossover first-order coupling rate* $\chi_1$: Inter-modal coupling through the two consecutive first-order diffraction happening once in the top grating layer and another in the bottom grating layer.

- *Straight-through second-order coupling rate* $\alpha_2$: Inter-modal coupling through the single second-order diffraction process in the entire grating layers.

See Fig. S1 for their graphical representations. We note that the crossover coupling appears only in the consecutive first-order diffraction processes in which the zero-order wave is involved as an intermediate state. In addition, imaginary parts in the coupling rates appear only in the first-order coupling rates $\alpha_1$ and $\chi_1$ for the same reason, *i.e.*, participation of the zero-order wave in the coupling processes as an intermediate state.

Looking further into the eigenvalue problem in Eq. (S13), the terms involving crossover coupling $\chi_1$ may vanish in certain conditions, resulting in further simplification of the problem. In fact, the crossover coupling must vanish in order to induce the critical topological phase which essentially requires zero net interaction between the left and right guided modes, *i.e.*, zero off-diagonal element in Eq. (S13). Therefore, existence of such conditions is important in the context of our study here. Such condition does exist and it is identical to the condition for the Fabry-Perot resonance of the zero-order wave. Combining Eq. (S16) for the $\chi_1$ definition and Eq. (S12) for the Green's function, we derive

$$\chi_1 = \omega_c \, \Delta\varepsilon_1^2 \, \frac{\zeta_0^{-1} k_c^2}{1-\rho^2} \left\{ \left[ e^{+iK_c h/2} + \rho e^{-iK_c h/2} \right] h \, \text{sinc}\left(\frac{K_c h}{2\pi}\right) \otimes U(K_c) \right\}^2, \tag{S18}$$

where symbol $\otimes$ denotes convolution with respect to argument $K_c$ and $U(K_c)$ is spatial Fourier transform of guided-mode wave function $u(z)$ which is assumed to be a symmetric real-valued function with respect to $z$. An immediate consequence of Eq. (S18) is

$$\chi_1 \approx 0 \text{ at waveguide thickness } d = 2h = 4\pi j \, K_c^{-1} \quad (j \text{ is a positive integer}) \tag{S19}$$

because the sinc function in Eq. (18) vanishes thereat. This condition is strictly valid for a guided mode loosely confined to the waveguide core as $U(K_c)$ appears as a narrow peak in $K_c$ domain and the convolution operation does not significantly alter the loci of the sinc function zeros. Even for tightly confined guided modes, one should be able to find $\chi_1 = 0$ at a thickness condition deviated from Eq. (S19) because the $U(K_c)$ peak cannot be significantly wider than $4\pi h^{-1}$ as a fundamental property of slab waveguide modes in general. We note in Eq. (S19) that the waveguide thickness condition is identical to the even-order Fabry-Perot resonance condition for the zero-order wave. This property further implies that the crossover zero-order coupling vanishes when the zero-order wave forms a standing wave in the waveguide layer with a symmetric field distribution with respect to $z$. Interestingly, the condition in (S19) is totally included in the antireflection condition of Eq. (4) in the main text because it is the Fabry-Pérot resonance condition including both even and odd orders. Therefore, the pure-phase resonance conditions of Eqs. (4) and (5) in the main text are favorably compatible with each other.

There exists another condition for $\chi_1 \approx 0$. It is the large fill-factor condition, *i.e.*, $F \approx 1$. In this limit, $\Delta\varepsilon_1^2 \approx (\varepsilon_H - \varepsilon_L)^2 (1-F)^2$ and thereby $\chi_1$ directly proportional to $\Delta\varepsilon_1^2$ quickly approaches 0 as $F$ approaches 1. On another hand, straight-through second-order coupling constant $\alpha_2$ is proportional to $\Delta\varepsilon_2 \approx (\varepsilon_H - \varepsilon_L)(1-F)$ for $F \approx 1$. Therefore, modal interaction in the large fill-factor limit is dominantly described by the terms associated with $\alpha_2$ while terms associated with $\chi_1$ yields negligible contribution. Importantly, the large-fill factor condition for $\chi_1 \approx 0$ is available for thin grating cases in which $d$ is substantially smaller than that for the lowest even-order Fabry-Pérot resonance condition in Eq. (S19). This is the case that we apply for the exemplary design in the main text.

For a structure at the optimal thickness condition in Eq. (S19), the eigenvalue problem in Eq. (S14) reduces to Eq. (1) in the main text with its parameters being determined by following relations.

$$\omega_0 = \omega_c + \text{Re}(\alpha_1), \qquad \gamma = \text{Im}(\alpha_1), \text{ and } m = \text{Re}(\alpha_1) - \alpha_2. \tag{S20}$$

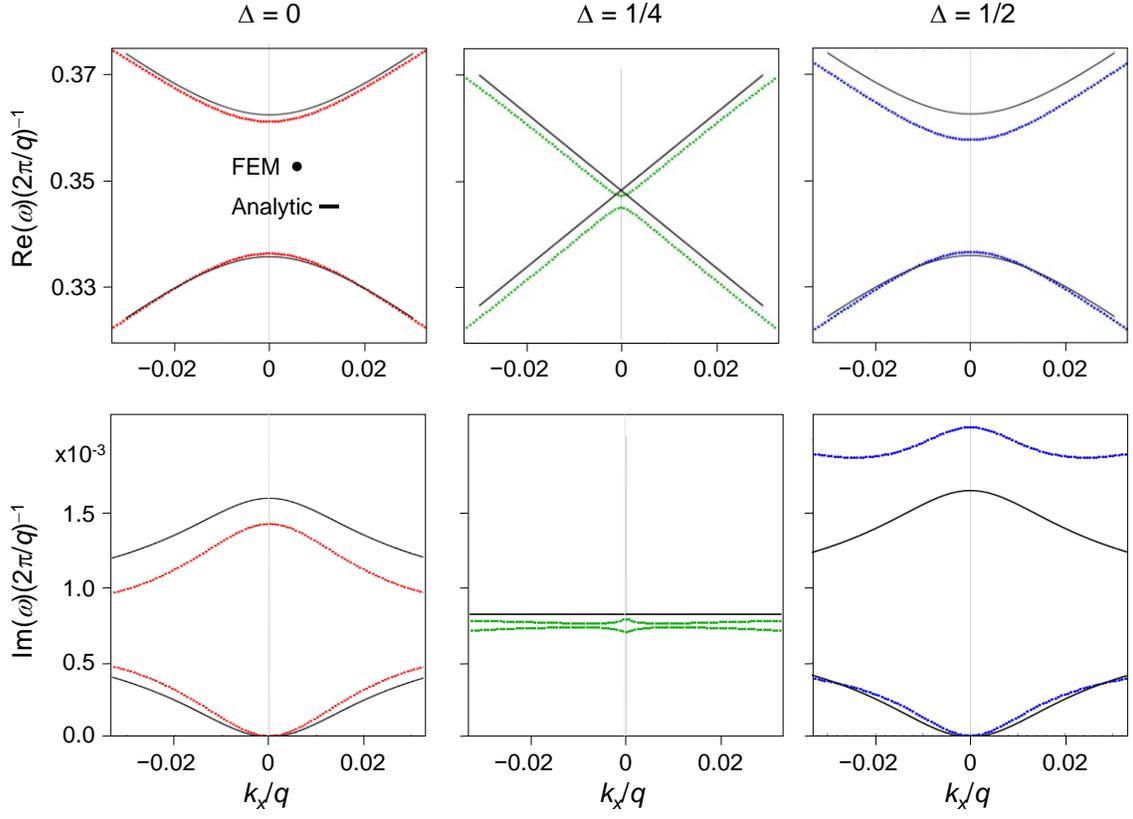

**Figure S2.** Real and imaginary band structures due to Eq. (S14) from our model in comparison with rigorous numerical calculation by the finite-element method. The structure is identical to that assumed in Figure 2 of the main text.

In Fig. S2, we show real and imaginary $\omega(k_x)$ bands due to Eq. (1) in comparison with rigorous numerical calculation by the finite-element method (FEM). We confirm that the complex frequency bands from this analytic model quantitatively captures major dependence of the band structures on shift parameter $\Delta$. In the analytic model calculation, we use Eqs. (S15)~(S17), and (S20) to determine the parameters in the Hamiltonian matrix.

S2. Leakage radiation properties

We explain basic properties of leakage radiation from resonantly excited guided-mode eigenstates in this supplementary section. In the absence of incident wave, leakage radiation $L(z)$ in Eq. (S5) takes an expression

$$L(z) = -S_{+1}(z)\psi_L - S_{-1}(z)\psi_R, \quad (S21)$$

where

$$S_{\pm 1}(z) = k_c^2 \int_{-h}^{+h} dz' \, G(z,z') \varepsilon_{\pm 1}(z') u(z'). \quad (S22)$$

Combining Eqs. (S21) and (S22) with the glide-symmetric dielectric function and the Green's function in Eqs. (S3), (S4), and (S12), we obtain more detailed expressions for the leakage radiation emitted towards the upper and lower background media as

$$L(z > +h) = L_c(z) = C_1 \left[ Ve^{+i\pi\Delta} + V^* e^{-i\pi\Delta} \, \psi_L + Ve^{-i\pi\Delta} + V^* e^{+i\pi\Delta} \, \psi_R \right] e^{+iK_d z}, \quad (S23)$$

$$L(z<-h)=L_{\text{s}}(z)=C_{1}\left[\left(V^{*}e^{+i\pi\Delta}+Ve^{-i\pi\Delta}\right)\psi_{\text{L}}+\left(V^{*}e^{-i\pi\Delta}+Ve^{+i\pi\Delta}\right)\psi_{\text{R}}\right]e^{-iK_{\text{d}}z}, \quad (S24)$$

where the coefficients are given by

$$C_{1}=-\Delta\varepsilon_{1}\frac{k_{\text{c}}^{2}}{iK_{\text{c}}}\frac{\tau e^{iK_{\text{d}}h}}{1-\rho^{2}}, \quad (S25)$$

$$V=V(k_{x},h)=\int_{0}^{h}dz'\left[e^{+iK_{\text{c}}z'}+\rho e^{-iK_{\text{c}}z'}\right]u(z'). \quad (S26)$$

Here, we assume that guided-mode wave function $u(z)$ is a real valued function of $z$. We note that the terms associated with $V$ is contributed from the upper grating and those associated with $V^*$ is contributed from the lower grating. Looking into the arithmetic form of Eqs. (S23) and (S24), we can further simplify the expressions for the leakage radiation as

$$L_{\text{c}}(z)=C_{1}|V|\left[\psi_{\text{L}}\cos(\pi\Delta+\phi)+\psi_{\text{R}}\cos(\pi\Delta-\phi)\right]e^{+iK_{\text{d}}z}, \quad (S27)$$

$$L_{\text{s}}(z)=C_{1}|V|\left[\psi_{\text{L}}\cos(\pi\Delta-\phi)+\psi_{\text{R}}\cos(\pi\Delta+\phi)\right]e^{-iK_{\text{d}}z}, \quad (S28)$$

where $\phi = \arg(V)$.

Although $\phi$ in general is a function of grating thickness $h$ and Bloch wavevector $k_x$, it has to take a value around $\phi=\pi/4$ when crossover coupling constant $\chi$ vanishes and the eigenvalue problem for the resonant mode subsequently reduces to Eq. (1) in the main text. This property is deduced from the radiation decay rate implied in the eigenvalue problem and its connection to the leakage-radiation wave functions in Eqs. (S27) and (S28).

First, for the resonances sufficiently away from Γ-point ($vk \gg m$), the radiation decay rate is given by

$$\begin{aligned}R_{\text{leak}}&=\text{Im}(\omega_{\pm})=\text{Im}\left[\omega_{0}+i\gamma\pm\sqrt{v^{2}k^{2}+(m+i\gamma)^{2}\cos^{2}(2\pi\Delta)}\right]\\ &\approx\gamma\pm\frac{m\gamma}{vk}\cos^{2}(2\pi\Delta)+O\left[\left(\frac{m\gamma}{vk}\right)^{2}\right]\approx\gamma\end{aligned}, \quad (S29)$$

because the terms involving "$m\gamma$" are second-order-perturbation quantities that we regard as negligibly contributing to the major interaction in our theory based on the first-order perturbation of diffractive processes. Therefore, we can consider that $R_{\text{leak}}$ is independent of shift parameter $\Delta$.

Next, we can alternatively derive an expression for $R_{\text{leak}}$ from the leakage-radiation wave functions in Eqs. (S27) and (S28) as

$$\begin{aligned}R_{\text{leak}}&=\eta\left[|L_{\text{cov}}|^{2}+|L_{\text{sub}}|^{2}\right]\\ &=2\eta|C_{1}V|^{2}\left\{\begin{array}{l}\left[\cos^{2}(\pi\Delta)\cos^{2}\phi+\sin^{2}(\pi\Delta)\sin^{2}\phi\right]\\ +2\text{Re}\left[\psi_{\text{L}}\psi_{\text{R}}^{*}\right]\cos(\pi\Delta+\phi)\cos(\pi\Delta-\phi)\end{array}\right\},\\ &\approx 2\eta|C_{1}V|^{2}\left[\cos^{2}(\pi\Delta)\cos^{2}\phi+\sin^{2}(\pi\Delta)\sin^{2}\phi\right]\end{aligned} \quad (S30)$$

where $\eta$ is a dimension-scaling constant and we include a relation $|\psi_{\text{L}}\psi_{\text{R}}| \ll 1$ for the condition away from Γ-point ($vk \gg m$) under the standard normalization scheme $|\psi_{\text{L}}|^2+|\psi_{\text{R}}|^2 = 1$. In Eq. (S30), the explicit dependence of $R_{\text{leak}}$ on shift parameter $\Delta$ appears and this $\Delta$ dependence must vanish in order to make the description consistent with Eq. (S29). The consistency between Eqs. (S29) and (S30) is obtained if and only if

$$\cos^{2}\phi=\sin^{2}\phi \text{ and } \therefore \phi=\pm\pi/4 \text{ or } \pm 3\pi/4. \quad (S31)$$

Proper choice of $\phi$ among four possible values depends dominantly on the thickness of the structure as implied in Eq. (S26). In our selected structure, the best fit to the simulated structure is obtained for $\phi = -\pi/4$ and this results in the leakage-radiation

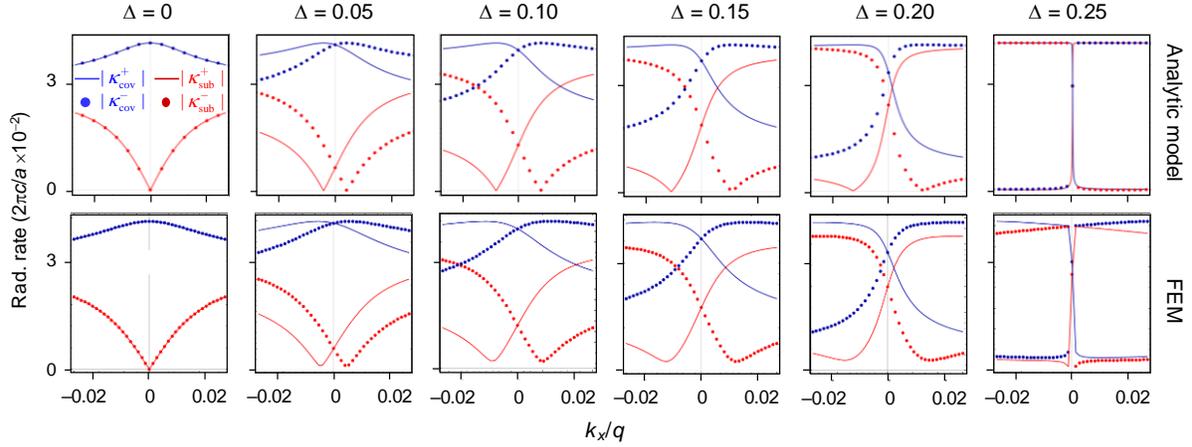

**Figure S3.** Leakage radiation strength due to the analytic model Eqs. (S34) and (S35) in comparison with the numerical simulation by the FEM method.

wave functions as

$$L_c(z) = C_1 |V| \left\{ \psi_L \cos[\pi(\Delta - 1/4)] - \psi_R \sin[\pi(\Delta - 1/4)] \right\} e^{+iK_d z}, \quad (S32)$$

$$L_s(z) = C_1 |V| \left\{ \psi_R \cos[\pi(\Delta - 1/4)] - \psi_L \sin[\pi(\Delta - 1/4)] \right\} e^{-iK_d z}, \quad (S33)$$

Equations (S32) and (S33) further allow closed form expressions of coupling constants $\kappa_c$ and $\kappa_s$ that describe emission of leakage radiation towards the cover and substrate from the resonant mode, respectively. Selecting the dimension of the coupling constants such that they naturally fit in the standard temporal coupled-mode theory for resonant scattering processes (24), the coupling constants follow scaling rule $|\kappa_c|^2 + |\kappa_s|^2 = 2R_{leak} = 2\gamma$ and their magnitudes are subsequently written by

$$|\kappa_c| = \sqrt{2\gamma} \left| \psi_L \cos[\pi(\Delta - 1/4)] - \psi_R \sin[\pi(\Delta - 1/4)] \right|, \quad (S34)$$

$$|\kappa_s| = \sqrt{2\gamma} \left| \psi_R \cos[\pi(\Delta - 1/4)] - \psi_L \sin[\pi(\Delta - 1/4)] \right|. \quad (S35)$$

We note here that Eqs. (S32)~(S35) describes the leakage-radiation waves respectively from $\psi_L$ and $\psi_R$ to be maximally asymmetric at $\Delta = \pm 1/4$ where the structure becomes maximally asymmetric in its geometry. In addition, these equations describes the leakage-radiation waves to be symmetric with respect to $\psi_L$ and $\psi_R$ at $\Delta = 0$ or $\pm 1/2$ where the structure becomes symmetric in its geometry. Therefore, Eqs. (S32) ~ (S35) provides physically reasonable descriptions as far as the symmetry in the structure geometry and scattered field distributions are taken into account.

In Fig. S3, we show $|\kappa_c|$ and $|\kappa_s|$ by Eqs. (S34) and (S35) in comparison with the rigorous numerical calculation due to the FEM method. We confirm quantitative agreement between two calculations. In this calculation, $\psi_L$ and $\psi_R$ are determined by

$$\begin{bmatrix} \psi_R^\pm \\ \psi_L^\pm \end{bmatrix} = N(k_x) \begin{bmatrix} \sqrt{\Delta\omega \pm v k_x} \\ \pm\sqrt{\Delta\omega \mp v k_x} \end{bmatrix}, \quad (S36)$$

where ($\pm$) sign distinguishes the upper (+) and lower (−) frequency eigenstates (the sign is taken in the same order), $N(k_x)$ is a normalization constant, and $\Delta\omega = [v^2 k_x^2 + (m + i\gamma)^2 \cos^2(2\pi\Delta)]^{1/2}$ is frequency spacing of eigenstates with respect to the bandgap center.

Describing leakage radiation properties, we define a leakage-radiation asymmetry factor

$$a_{cs} = \left| \frac{\kappa_c}{\kappa_s} \right| = \left| \frac{\psi_L \cos[\pi(\Delta - 1/4)] - \psi_R \sin[\pi(\Delta - 1/4)]}{\psi_R \cos[\pi(\Delta - 1/4)] - \psi_L \sin[\pi(\Delta - 1/4)]} \right|. \quad (S37)$$

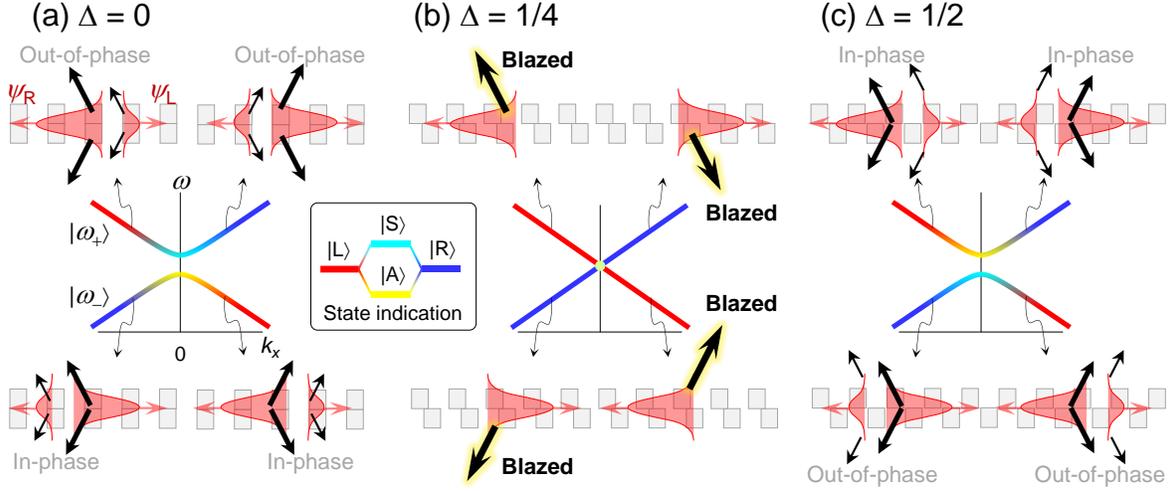

**Figure S4.** Leakage radiation properties for different glide-shift parameters. (a) Δ = 0 (trivial mirror-symmetric structure). (b) Δ = 1/4 (maximally asymmetric structure and nodal phase condition). (c) Δ = 1/2 (another trivial mirror-symmetric structure).

$a_{cs}$ indicates how much the leakage radiation $L_{cov}$ towards the cover medium is stronger than $L_{sub}$ towards the substrate medium and Eq. (S37) conveniently describes it as a function of the left and right guided mode amplitudes and shift parameter Δ. Blazing of the leakage radiation is readily understood with Eq. (S37). We consider a case for $k_x > 0$, sufficiently away from Γ point for example. We can estimate the guided-mode amplitudes $\psi_L \approx 0$ and $\psi_R \approx 1$. Then, $a_{cs} \approx |\tan[\pi(\Delta - 1/4)]|$. This results in $a_{cs} \approx 1$ for Δ = 0 or 1/2, *i.e.*, the leakage radiation is uniformly emitted towards the cover and substrate if the structure is mirror symmetric with respect to *x-y* plane. In addition, $a_{cs} \approx 0$ or ∞ for Δ = +1/4 or –1/4, *i.e.*, the perfect leakage-radiation blazing is obtained if the structure is maximally asymmetric with respect to *x-y* plane.

In another consideration, the glide symmetry of the structure implies associated symmetry properties in the resonant guided modes such that

$$\omega(\pm k_x) = \omega(\mp k_x), \tag{S38}$$

$$\psi_L(\pm k_x) = \psi_R(\mp k_x). \tag{S39}$$

These properties further imply that the leakage radiation distribution follows subsequent symmetry properties as

$$\kappa_c(\pm k_x) = \kappa_s(\mp k_x), \tag{S40}$$

$$a_{cs}(\pm k_x) = 1/a_{cs}(\mp k_x). \tag{S41}$$

These symmetry relations describe the resonant modes and leakage radiation distributions as summarized in Fig. (S4). The temporal coupled-mode theory in the next section is based on these symmetry relations.

S3. Temporal coupled-mode theory for the spectral properties

In this section, we develop a model for the transmission and reflection coefficients based on the temporal coupled-mode theory. We use a four-port two-state system to conveniently describe our problem of a glide-symmetric bi-layer waveguide grating structure, as schematically illustrated in Fig. S4. The frequency-domain coupled mode equations for this system are written as

$$-i\omega |U\rangle_G = -i(\boldsymbol{\omega}_0 - i\boldsymbol{\Gamma})|U\rangle_G + (\boldsymbol{\kappa}\boldsymbol{\sigma}_x)^T |A\rangle_R, \tag{S42}$$

$$|B\rangle_R = \mathbf{C}|A\rangle_R + \boldsymbol{\kappa}|U\rangle_G, \tag{S43}$$

where resonant mode state vector $|U\rangle_G$ in the guided-mode field, incoming wave and outgoing wave state vectors $|A\rangle_R$ and $|B\rangle_R$ in the radiation-continuum field are respectively defined by

$$|U\rangle_G = \begin{bmatrix} U_+ \\ U_- \end{bmatrix}, \quad |A\rangle_R = \begin{bmatrix} A_1 \\ A_2 \\ A_3 \\ A_4 \end{bmatrix}, \text{ and } |B\rangle_R = \begin{bmatrix} B_1 \\ B_2 \\ B_3 \\ B_4 \end{bmatrix}, \tag{S44}$$

resonance frequency matrices

$$\boldsymbol{\omega}_0 = \begin{bmatrix} \text{Re}[\omega(+k_x)] & 0 \\ 0 & \text{Re}[\omega(-k_x)] \end{bmatrix} \text{ and } \boldsymbol{\Gamma} = \begin{bmatrix} \text{Im}[\omega(+k_x)] & 0 \\ 0 & \text{Im}[\omega(-k_x)] \end{bmatrix}, \tag{S45}$$

non-resonant scattering and resonant coupling matrices

$$\mathbf{C} = \begin{bmatrix} 0 & -r & 0 & it \\ -r & 0 & it & 0 \\ 0 & it & 0 & -r \\ it & 0 & -r & 0 \end{bmatrix} \text{ and } \boldsymbol{\kappa} = \begin{bmatrix} 0 & \kappa_1 \\ \kappa_2 & 0 \\ 0 & \kappa_3 \\ \kappa_4 & 0 \end{bmatrix} \tag{S46}$$

with magnitudes of the coupling matrix elements being determined according to Eqs. (S34) and (35) such that

$$|\kappa_1| = |\kappa_{\text{cov}}(-k_x)|, \ |\kappa_2| = |\kappa_{\text{cov}}(+k_x)|, \ |\kappa_3| = |\kappa_{\text{sub}}(-k_x)|, \text{ and } |\kappa_4| = |\kappa_{\text{sub}}(+k_x)|. \tag{S47}$$

Here, we note that the glide symmetry of the structure implies Eq. (S40) and thereby

$$\kappa_1 = \kappa_4 \text{ and } \kappa_2 = \kappa_3. \tag{S48}$$

For this symmetry property, the resonant coupling matrix in (S46) reduces to

$$\boldsymbol{\kappa} = \begin{bmatrix} 0 & \kappa_1 \\ \kappa_2 & 0 \\ 0 & \kappa_2 \\ \kappa_1 & 0 \end{bmatrix} \tag{S49}$$

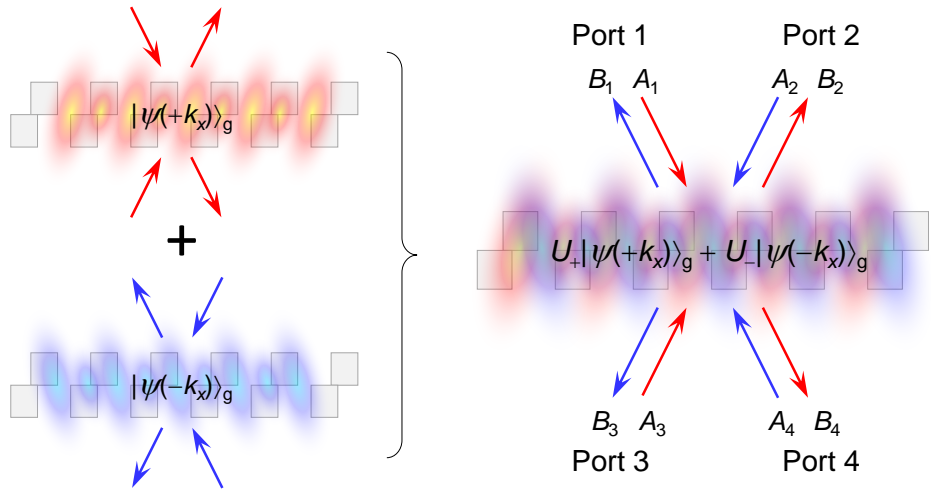

**Figure S5.** Localized state and radiation mode configuration for the temporal coupled-mode theory construction. We regard the system to be a four-port two-state system as a superposition of total field configurations for in-plane wave vectors $+k_x$ and $-k_x$ for complete description of spatial symmetries implied in the structure and its response.

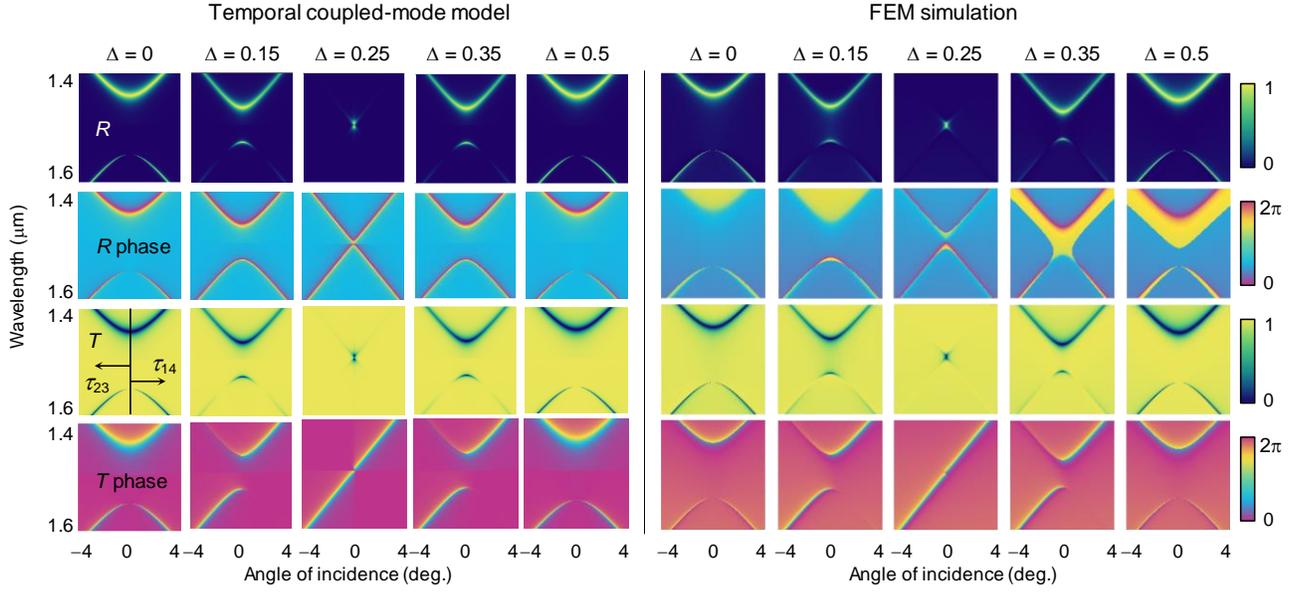

**Figure S6.** Intensity and phase spectra of the reflection and transmission calculated by the temporal coupled-mode model and the finite-element method simulation.

with magnitudes of $\kappa_1$ and $\kappa_2$ being related to each other by radiation asymmetry factor

$$\alpha = \left|\frac{\kappa_2}{\kappa_1}\right| = a_{cs}(+k_x) = \frac{1}{a_{cs}(-k_x)}. \tag{S50}$$

Here, $a_{cs}(k_x)$ is given by Eq. (S37). With these parametric setting, energy conservation requirement $\boldsymbol{\kappa}^\dagger\boldsymbol{\kappa} = 2\boldsymbol{\Gamma}$ is naturally satisfied.

Imposing the time-reversal symmetry of the scattering processes (*24*), resonant coupling matrix $\boldsymbol{\kappa}$ and non-resonant scattering matrix **C** are related to each other by constraints

$$\mathbf{C}\boldsymbol{\kappa}^* = -\boldsymbol{\kappa}, \tag{S51}$$

and it results in the determination of resonant coupling phase difference $\Delta\phi = \arg(\kappa_2) - \arg(\kappa_1)$ as a function of $a$, $r$, and $t$ as

$$\Delta\phi = \sin^{-1}\left[\frac{(\alpha^2 - 1)r}{2\alpha t}\right]. \tag{S52}$$

Equation (S52) yields useful expressions for the resonant coupling constants as functions of phase difference $\Delta\phi$ and non-resonant scattering amplitudes $r$ and $t$ as

$$\kappa_1 = \sqrt{\frac{2\gamma}{1+\alpha^2}} \exp\left[i\frac{1}{2}\arg\left(it + \alpha r e^{-i\Delta\phi}\right)\right], \tag{S53}$$

$$\kappa_2 = \sqrt{\frac{2\gamma}{1+\alpha^{-2}}} \exp\left[i\frac{1}{2}\arg\left(it + \alpha^{-1} r e^{-i\Delta\phi}\right)\right]. \tag{S54}$$

Including Eqs. (S53) and (S54) in coupled-mode Eqs. (S42) and (S43), we finally obtain a scattering matrix equation as

$$|B\rangle_R = \mathbf{S}|A\rangle_R = \left\{\mathbf{C} + \boldsymbol{\kappa}\left[i(\omega - \omega_0) - \boldsymbol{\Gamma}\right]^{-1}(\boldsymbol{\kappa}\boldsymbol{\sigma}_x)^{\mathrm{T}}\right\}|A\rangle_R, \tag{S55}$$

and the scattering matrix elements $S_{nm}$ herein take closed-form expressions as

$$S_{12} = S_{21} = S_{34} = S_{43} = -r + \frac{\kappa_1 \kappa_2}{i(\omega - \omega_0) - \Gamma} \equiv \rho, \tag{S56}$$

$$S_{14} = S_{41} = it + \frac{\kappa_1^2}{i(\omega - \omega_0) - \Gamma} \equiv \tau_{14}, \tag{S57}$$

$$S_{23} = S_{32} = it + \frac{\kappa_2^2}{i(\omega - \omega_0) - \Gamma} \equiv \tau_{23}, \tag{S58}$$

with all the other elements being zero. Here, $\Gamma = \text{Im}[\omega(+k_x)] = \text{Im}[\omega(-k_x)]$, $\rho$ is reflection coefficient common at all radiation ports, $\tau_{14}$ is transmission coefficient between ports 1 and 4, and $\tau_{23}$ is transmission coefficient between ports 2 and 3. In Fig. S6, we show reflectance $R = |\rho|^2$, reflection phase $\phi_r = \arg(\rho)$, transmittance $T = |\tau_{nm}|^2$, and transmission phase $\phi_t = \arg(\tau_{nm})$ spectra due to Eqs. (S56)~(S57) in comparison with the FEM simulation. We confirm quantitative agreement of our model with the rigorous numerical calculation.